\documentclass[aps,prb,twocolumn,superscriptaddress]{revtex4-1}

\usepackage{xcolor}
\usepackage{soul}
\usepackage{enumitem}
\usepackage{multirow}
\usepackage{amsmath}
\usepackage{graphicx}
\usepackage{color}
\usepackage{combelow}

\begin{document}

\title{Enhancing the sensitivity and selectivity of pyrene-based sensors for 
detection of small gaseous molecules via  destructive quantum interference}

\author{Ozlem Sengul}
\affiliation{Institute for Theoretical Physics, Vienna University of Technology, Wiedner Hauptstrasse 8-10, 1040 Vienna, Austria}
\author{Julia V\"olkle}
\affiliation{Department of Physical Chemistry, University of Vienna, W\"ahringer Strasse 42, 1090 Vienna, Austria}  
\affiliation{Centre of Electrochemical Surface Technology, Viktor Kaplan-Strasse 2, 2700 Wiener Neustadt, Austria}
\author{Angelo Valli}
\affiliation{Institute for Theoretical Physics, Vienna University of Technology, Wiedner Hauptstrasse 8-10, 1040 Vienna, Austria}
\author{Robert Stadler}
\affiliation{Institute for Theoretical Physics, Vienna University of Technology, Wiedner Hauptstrasse 8-10, 1040 Vienna, Austria}

\begin{abstract} 
Graphene-based sensors are exceptionally sensitive with high carrier mobility and low intrinsic noise, and have been intensively investigated in the past decade. The detection of individual gas molecules has been reported, albeit the underlying sensing mechanism is not yet well understood. 
We focus on the adsorption of NO$_2$, H$_2$O, and NH$_3$ on a molecular junction with a pyrene core, which can be considered as a minimal graphene-like unit. 
We systematically investigate the chemiresistive response within the framework of density functional theory and non-equilibrium Green’s functions.
We highlight the fundamental role of quantum interference (QI) in the sensing process, and we propose it as a paradigmatic mechanism for sensing. 
Owing to the open-shell character of NO$_2$, its interaction with pyrene gives rise to a Fano resonance thereby triggering the strongest chemiresistive response, while the weaker interactions with H$_2$O and NH$_3$ result in lower sensitivity. We demonstrate that by exploiting destructive QI arising in the meta-substituted pyrene, it is possible to calibrate the sensor to enhance both its sensitivity and chemical selectivity by almost two orders of magnitude so that individual molecules can be detected and distinguished. 
These results provide a fundamental strategy to design high-performance chemical sensors with graphene functional blocks.
\end{abstract}

\maketitle

\section*{Introduction}

Gas sensing technologies are of paramount importance for environmental safety~\cite{buckley2020frontiers} and medical applications.~\cite{nasiri2019nanostructured} 
Solid-state sensors are prominent with their achieved sensitivity, in addition to low cost, and miniature size, 
making them widely used in several applications.~\cite{moseley1997solid,azad1992solid,capone2003solid} 
Among carbon-based materials, carbon nanotubes have been reported to have a fast response and high  sensitivity.~\cite{kong2000nanotube,wang2009review} 
It has also been stated in the context of gas sensors based on thin films of organic polymers 
that a monolayer thick semiconductor yields the ultimate sensing performance.~\cite{andringa2010pa} 
Single-layer graphene naturally fulfills this requirement. 
The unique characteristics of graphene indicate its major potential for applications in nanoelectronics.~\cite{westervelt2008graphene,sato2015graphene} 
Moreover, due to the very large changes in conductivity that are possible 
as a result of charge doping of graphene sheets with a number of adsorbates,~\cite{kong2014molecular} 
together with its uniquely high  surface-to-volume ratio and low electrical noise, 
the material is a promising candidate for next-generation sensors with performances beyond the reach of solid-state devices. 
Indeed, graphene has been proposed for gas sensing,~\cite{ratinac2010toward,wang2016review} as a bio sensor,~\cite{reiner2015graphene} 
and even DNA base sequence analysis.~\cite{merchant2010dna,saha2012dna} 
Recent concepts for gas sensing based on graphene exploit its mechanical properties as a resonating membrane.~\cite{dolleman2016graphene,dolleman2016} 
The chemiresistor concept~\cite{frazier2013robust} nevertheless remains the most widely used in gas/vapor sensors, 
where a voltage is applied on two electrodes connected to a graphene sheet and the resulting current fluctuations 
in dependence on the gas composition are recorded.~\cite{wang2016review} 
The same concept is also used in bio-sensors, where graphene needs to be chemically functionalized with an antibody~\cite{reiner2015graphene} 
in order to achieve biochemical selectivity to a specific antigen 
but where non-specific binding or interaction of other solute molecules directly with the graphene sheet remain a severe problem.
Already a decade ago the detection of a single NO$_2$ molecule with a graphene based sensor was reported,~\cite{schedin2007detection} 
and the sensitivity of graphene-based sensors has been increased in recent experiments, 
in particular for H$_2$O~\cite{smith2015resistive} and NO$_2$.\cite{seekaew2017highly}
The reviews on gas sensors based on graphene~\cite{kong2014molecular,ratinac2010toward,wang2016review} 
identify the same two major challenges for their further development: 
(i) the mechanism on which the gas sensing is based is not very well understood and sometimes several competing theoretical explanations exist;~\cite{ratinac2010toward} 
(ii) while there is little doubt that thin graphene films show great sensitivity, 
unfortunately they are sensitive to many different types of adsorbates and this cross-sensitivity naturally diminishes 
another important property of any chemical sensor, namely chemical selectivity.

These problems can potentially be overcome by replacing extended graphene sheets with graphene nanostructures, 
such as atomically well-defined graphene nanoribbons (GNRs), 
for which new paradigms for the detection (or switching) mechanism need to be proposed. 
Tremendous progress has been made recently in the chemical synthesis 
of atomically precise~\cite{cai2010atomically,chen2015molecular,elabbassi2019robust,caiyao2020fabrication,prins2011room} 
and liquid-phase-processable~\cite{narita2014synthesis} GNRs, 
which can be functionalized for sensing applications~\cite{wei2012enhanced,alzate2020functionalized} 
and easily integrated in junctions or electric circuits.~\cite{caneva2018mechanically} 
At the same time, quantum interference (QI) effects, which are well established in $\pi$-conjugated single-molecule junctions,~\cite{tsuji2018quantum,markussen2010electrochemical, markussen2011graphical,stadler2011controlling,zhao2017quantum,stadler2003modulation,stadler2004integrating,stadler2009quantum,zhao2017destructive,zhao2019dft,stadler2005ab} 
have also been theoretically predicted~\cite{munarriz2011towards,sangtarash2016exploring,valli2018quantum,valli2019interplay,nictua2020robust} 
and experimentally observed~\cite{gehring2016quantum,caneva2018mechanically} in nanostructured graphene. 
In terms of applications, QI has also been suggested as a new paradigm 
for logic devices based on GNRs with extremely low power consumption~\cite{gehring2016quantum,sadeghi2015conductance} 
as well as a tool for increasing the selectivity of GNR-based gas sensors.~\cite{wei2012enhanced} 

We propose the concept of a high-performance chemical sensor operating with QI effects, and we demonstrate the detection of individual molecular adsorbates. 
In this study, we focus on a disubstituted pyrene molecule bridging Au electrodes in a single-molecule junction setup. For the purpose of gas sensing, we investigate the adsorption of NO$_2$, H$_2$O, and NH$_3$ on the pyrene core. Understanding the physical, chemical, and electronic interactions between the pyrene and the adsorbate is essential to gain insight into the sensing mechanism. We show that QI play a fundamental role in determining the chemiresistive response, and we propose a protocol to enhance both the sensitivity and the chemical selectivity to the molecular dopant. 
The QI properties of the junction are entirely determined by the contact configuration of the pyrene core,~\cite{sengul2021electrode} 
which, as a polycyclic aromatic hydrocarbon, can be considered a minimal graphene-like  molecule.~\cite{sangtarash2015searching,sangtarash2016exploring,sengul2021electrode} 
The occurence of QI in alternant hydrocarbons do not depend on the precise shape 
of the molecular bridge but arise from the sublattice  structure.~\cite{zhao2017destructive,valli2018quantum,valli2019interplay,nictua2020robust,sengul2021electrode} 
Hence, the concept of QI-enhanced chemiresistor proposed here 
is general enough to be extended to different structures, 
including other molecules as well as graphene nanoribbons, 
and can thus provide new design strategies for nanosensors.



\begin{figure}[bp]
    \begin{center}
        \includegraphics[width=0.47\textwidth]{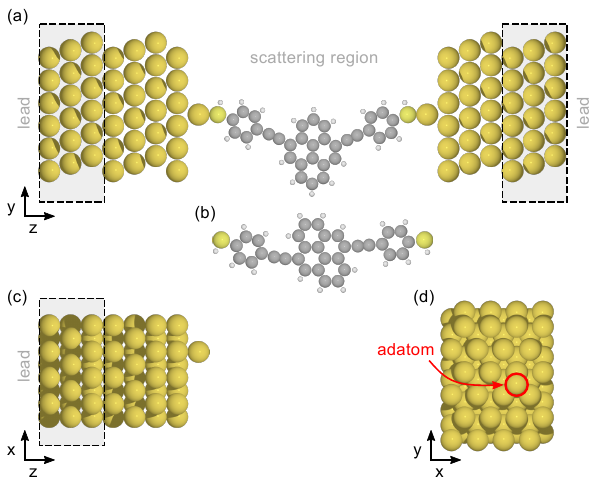}
    \end{center}
    \caption{(a) Structure of the meta-substituted pyrene single-molecule junction, connected to the Au electrodes via propynylbenzene linkers terminated with thiol anchor groups. 
    Both the periodic Au lead unit cell and the scattering region are shown. 
    (b) Para-substituted pyrene molecule. 
    (c,d) Transverse sections of the Au electrodes in the $xz$ and $xy$ planes. 
    The adatom is in the hollow position of the Au(111) surface.}
    \label{fig:structure}
\end{figure} 

\section*{Computational details}

The adsorption of the selected molecules on the pyrene core and the chemiresistive response are investigated 
within the framework of density functional theory and non-equilibrium Green's functions,\cite{brandbyge2002DFTNE,xue2002first} (DFT+NEGF)
as implemented in the Atomic Simulation Environment~\cite{larsen2017ASE} (ASE) and the GPAW software packages.~\cite{mortensen2005GPAW,enkovaara2010GPAW} 
The electron wave functions are described by atom-centered basis functions within an LCAO double-$\zeta$ polarized basis set, 
with a grid spacing of 0.2~\AA, and we employ the Perdew-Burke-Ernzerhof (PBE) parametrization for the exchange-correlation functional. 


\begin{figure}[tbp]
    \begin{center}
        \includegraphics[width=0.45\textwidth]{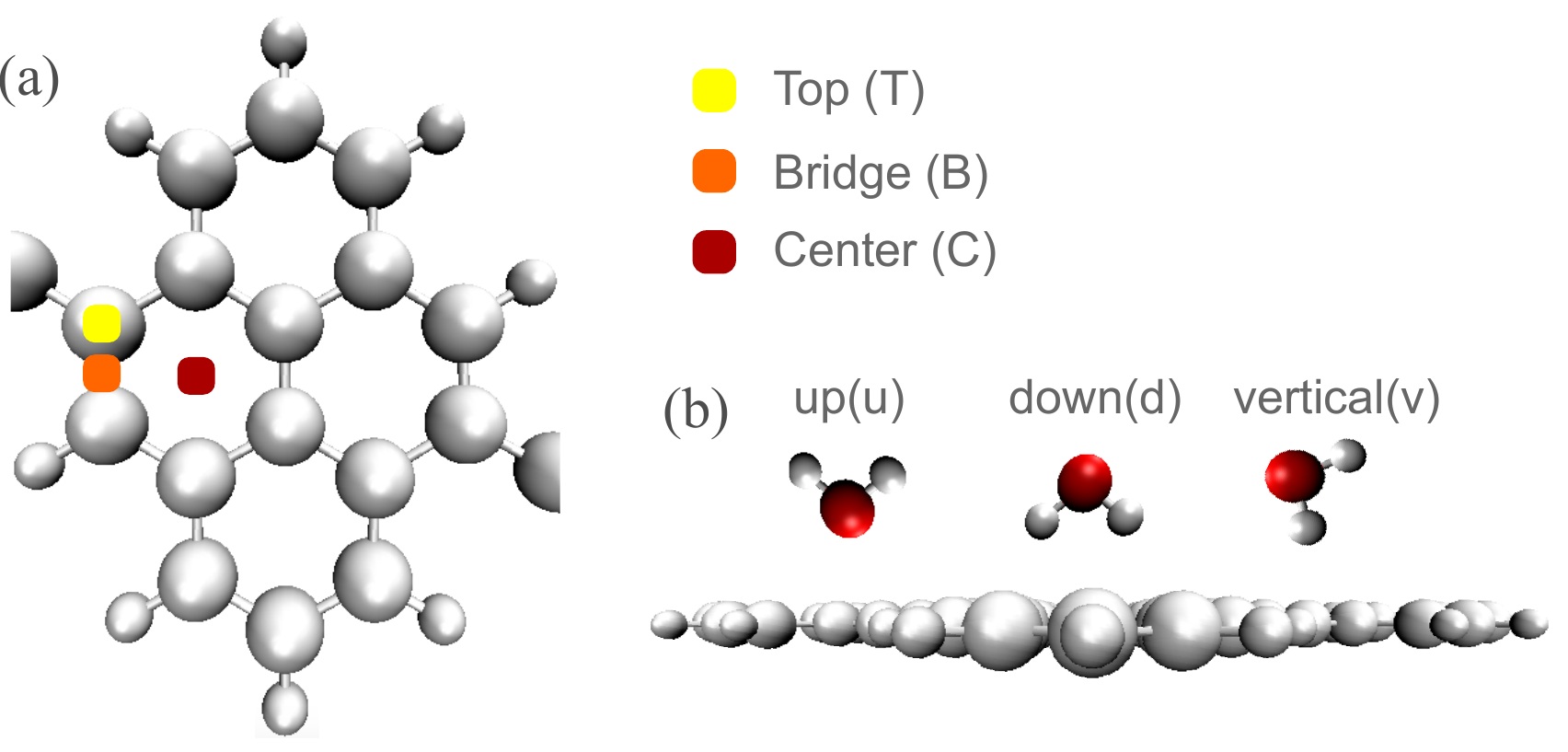}
    \end{center}
    \caption{(a) Adsorption sites on pyrene and (b) orientations chosen for the placement of the adsorbates, shown for the case of H$_2$O.}
    \label{fig:position}
\end{figure}

\begin{figure*}[t]
    \centering
    \includegraphics[width=1.0\textwidth]{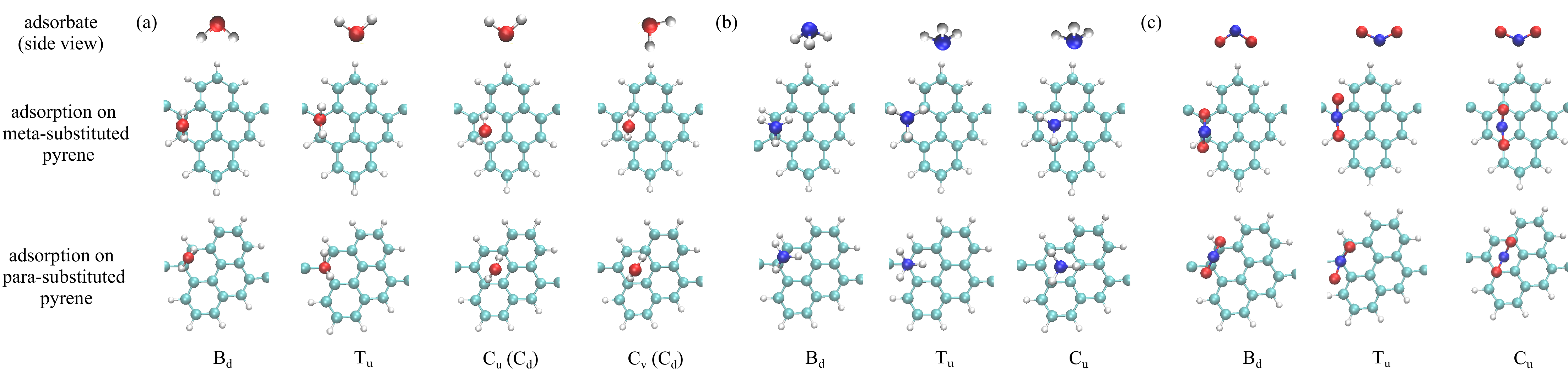}
    \caption{Adsorption of (a) H$_2$O, (b) NH$_3$, and (c) NO$_2$ 
    on meta- and para-substituted pyrene cores. 
    [Top panels] Orientation of the adsorbate before the structural relaxation (side view). 
    [Lower panels] Adsorption position on pyrene and orientation of the adsorbate 
    after the structural relaxation (top view). 
    In the case of H$_2$O, the initial C$_u$ and C$_v$ orientations are not stable 
    and the adsorbate prefers the C$_d$ orientation after relaxation (as indicated in brackets). }
    \label{fig:pyrene-positions}
\end{figure*}

In the following, we consider molecular junctions where a pyrene core  
is connected to Au electrodes via propynylbenzene linkers terminated with thiol anchor groups, 
in two different configurations. 
We denote a configuration as \emph{meta} or \emph{para}, depending whether it exhibits 
destructive (DQI) or constructive (CQI) interference.~\cite{sengul2021electrode} 
A detailed representation of the single-molecule junctions is shown Fig.~\ref{fig:structure}.
The adsorbates are placed in the vicinity of the pyrene carbon atom, which bonds to the left linker, 
either on top of the carbon atom (T configuration), 
on the bond with the neighbouring carbon atom (B configuration), 
or in the center of the corresponding benzene ring (C configuration). 
Besides different placements, we also consider different orientations of the adsorbates, 
which can either face with its ``legs" (i.e., the hydrogen or, in the case of NO$_2$, the oxygen atoms) 
away from (up configuration, $u$) or towards (down configuration, $d$) the pyrene core. 
In the case of H$_2$O, we also consider a specific configuration (vertical, $v$) in which the molecule is oriented with one leg towards the junction, 
while the other one is parallel to the pyrene plane. 
Those orientations have been identified as the most stable ones for adsorption on graphene.~\cite{leenaerts2008adsorption,hamada2012adsorption,leenaerts2009water,leenaerts2009adsorption} 
An overview of the adsorption positions and orientations is shown in Figs.~\ref{fig:position} and \ref{fig:pyrene-positions}. 
For each configuration, we perform a structural optimization of the pyrene molecule (including the linkers) and the adsorbate in gas phase 
until the Hellmann-Feynman forces were below $0.05$~eV/\AA. 
The equilibrium distance is defined as the distance between the pyrene adsorption position and the N or O atoms of H$_2$O, NH$_3$, and NO$_2$, 
and the adsorption energy E$_a$ is calculated as follows: 
\begin{equation} \label{eq:Ea}
    {E_a} = E_{\mathrm{pyrene+adsorbate}} - E_{\mathrm{pyrene}} - E_{\mathrm{adsorbate}},
\end{equation} 
where each term is evaluated in the relaxed atomic structure. 

For the electrodes, the scattering region consists of seven $6\times 4$-layers Au(111), 
and the molecular bridge is connected via thiol anchoring groups to Au adatoms placed in the hollow position of the Au(111) surface. 
We take a typical bonding distance of $d_{\textrm{Au-S}}=2.12$~\AA.~\cite{zhao2017destructive,stadler2005forces} 
For the electron transport calculations, the scattering region is sampled with a $4 \times 4 \times 1$ Monkhorst-Pack mesh, 
and the leads are sampled with a $4 \times 4 \times 6$ Monkhorst-Pack mesh, where $z$ denotes the transport direction. 
The transmission function is calculated within the Landauer-B\"{u}ttiker formalism~\cite{landauer1957spatial} as 
\begin{equation}
 T(E) = \mathrm{Tr}\Big[ \Gamma^L(E) G^{\dagger}(E) \Gamma^R(E) G(E) \Big],
\end{equation}
where $G$ is the retarded Green's function of the scattering region 
and $\Gamma = \imath(\Sigma-\Sigma^{\dagger})/2$ denotes the coupling to either the left (L) or right (R) lead in terms of the corresponding embedding self-energy.  
Finally, we evaluate the electric current (per spin) as
\begin{equation}
    I = \frac{e}{h} \int_{-\infty}^{\infty} dE \ T(E) \ \big[ f_S(E) - f_D(E) \big],
\end{equation}
where $e$ is the electric charge and $h$ the Planck constant, 
while the Fermi distribution function of the source (S) and drain (D) electrodes is given by
\begin{equation}
 f_{S/D}(E) = \frac{1}{1 + \exp[(E - V_{S/D})/k_BT]}, 
\end{equation}
where $k_B$ is the Boltzmann constant and $V_b=V_{S}-V_{D}$ is the symmetric bias drop 
between the source and the drain.

\section*{Results and Discussion}
Our discussion is structured as follows. First, we characterize the adsorption 
of the gaseous molecules on the pyrene core in terms of adsorption energy and distance 
in order to identify the most stable configuration for each adsorbate. 
For those, we analyze the sensitivity and selectivity of the junction, 
focusing on experimentally accessible quantities, 
such as the zero-bias conductance and the I-V characteristics. 
Finally, we understand the electronic response of the system 
to the presence of each adsorbate 
via a transmission function and molecular orbital (MO) analysis, 
and we highlight the role of QI effects in enhancing the sensing capability 
of the device. 

Before turning to the results, it is useful to put our analysis in a general perspective. 
It is inevitable that a quantitative characterization of the adsorption shall depend 
on the technical details of the methodology employed. 
In particular, the relaxed atomic configuration and the electronic properties 
(including the adsorption energy) depend on the chosen exchange energy functional, 
and whether other effects, such as van der Waals interactions, are included. 
In the literature, this has been explicitly discussed 
for the adsorption of H$_2$O and other polar molecules on graphene, 
and it was found that, in general, the adsorption energy might change substantially 
due to van der Waals contributions.~\cite{hamada2012adsorption,lin2013adsorption} 
At the same time, the HOMO-LUMO gap can be corrected 
with more or less sophisticated approaches 
that take into account many-body effects beyond density functional theory. 
The position of the transmission resonances affects the value of the conductance 
and the evaluated sensitivity may not accurately reproduce experimental results. 

However, the QI properties of the junctions discussed here, 
i.e., (i) the emergence of a Fano resonance due to the adsorption of NO$_2$ 
and (ii) the occurrence of a DQI antiresonance in the meta-substituted pyrene, 
are qualitatively robust due to their specific physical origin. 
An indirect evidence of this stability is that QI effects are routinely 
observed in the experiments, despite the statistical nature 
of the transport measurements in a break-junction setup. 

Eventually, the scope of our analysis is 
to propose the concept of a QI-enhanced chemiresistive sensor, 
and we focus on a few cases to demonstrate how it works, 
transcending the intrinsic technical uncertainties of the numerical simulations.

{\renewcommand{\arraystretch}{0.5}
\begin{table}[bp]
\centering
\small
  \caption{Adsorption energies and the distance of all selected adsorbates. 
  The most stable configuration for each adsorbate is highlighted. 
  For each entry, we indicate the initial configuration, i.e., before the structural relaxation, 
  and the relaxed one in brackets, if different (see text for a discussion). }
  \label{table:Ad}
  \begin{tabular*}{0.45\textwidth}{@{\extracolsep{\fill}}clcccc}
    \hline
    & & \multicolumn{2}{c}{meta} & \multicolumn{2}{c}{para} \\
    \cline{3-4} \cline{5-6}
    & & E$_{a}$(eV) & d(\AA) & E$_{a}$(eV) & d(\AA)    \\
    \hline
    H$_2$O    & B$_d$ & -0.181 & \phantom{-}3.2  & -0.188 & 3.3 \\
              & T$_u$ & -0.050 & \phantom{-}3.1  & -0.082 & 3.1 \\
              & C$_u$ (C$_d$) & -0.195 & \phantom{-}3.1  & -0.196 & 3.2 \\
              & \textbf{C$_v$} (C$_d$) & \textbf{-0.198} & \textbf{\phantom{-}3.3} & \textbf{-0.198} & \textbf{3.2} \\ 
    NH$_3$    & \textbf{B$_d$}& \textbf{-0.147}  & \textbf{\phantom{-}3.3} & \textbf{-0.131} & \textbf{3.4} \\
              & T$_u$ & -0.038 & \phantom{-}3.2  & -0.027 & 3.0\\
              & C$_u$ & -0.048 & \phantom{-}3.0  & -0.046 & 3.0 \\  
    NO$_2$    & \textbf{B$_d$} & \textbf{-0.257} & \textbf{\phantom{-}3.4} & \textbf{-0.198} & \textbf{3.4} \\
              & T$_u$ & -0.167 & \phantom{-}3.4  & -0.163 & 3.2 \\
              & C$_u$ & -0.193 & \phantom{-}2.7  & -0.170 & 3.1 \\
    \hline
  \end{tabular*}
\end{table}
}

{\renewcommand{\arraystretch}{0.5}
\begin{table}[btp]
\centering
\small
  \caption{Bader charges of the adsorbates ($\Delta{Q}$) in the most stable configurations (among those of Table~\ref{table:Ad}) on meta- and para-substituted pyrene.}
  \label{table:Charge}
  \begin{tabular*}{0.30\textwidth}{@{\extracolsep{\fill}}clcc}
    \hline
    & & meta           & para           \\ 
    & & $\Delta{Q}$($|e|$) & $\Delta{Q}$($|e|$) \\
    \hline
    H$_2$O   
              & C$_v$ (C$_d$) & -0.021 & \phantom{-}-0.022 \\ 
    NH$_3$    & B$_d$ & -0.011 & \phantom{-}-0.009 \\
             
    NO$_2$    & B$_d$ & -0.192 & \phantom{-}-0.161 \\
           
    \hline
  \end{tabular*}
\end{table}
}

\subsection*{Adsorption distances and energies}
The adsorption energies and distances for the selected adsorbates are reported in Table~\ref{table:Ad}. 
All the calculated adsorption energies are found to be negative,
ranging between $-0.1$ and $-0.3$~eV. Both the adsorption energy and the adsorption distance (around $3$~\AA) 
indicate that all molecules are physically adsorbed, in agreement with the results of adsorption on graphene 
in the literature.~\cite{leenaerts2008adsorption,lin2013adsorption,zhang2009improving,bottcher2011graphene,kumar2018adsorption} 

Comparing the adsorption energies, we find the most stable configurations to be B$_d$ for NO$_2$ and NH$_3$, and C$_v$ for H$_2$O. 
Note, however, that during the relaxation process of both the C$_u$ and the C$_v$ configurations, 
the H$_2$O molecule rotates so that in the final configurations, the H atoms point towards the surface (C$_d$ configuration). 
As a result, the adsorption energies of the two final configurations are very similar. 
The C$_d$ configuration was found to be the most stable for H$_2$O on graphene 
even when van der Waals interactions are taken into account,~\cite{lin2013adsorption} 
but in general the adsorption energy might change substantially due to van der Waals contributions.~\cite{hamada2012adsorption,lin2013adsorption}

The preferred orientations can be understood as follows. 
The electron-rich C atom in the pyrene molecule tends to form a $\pi$-hydrogen bond with X-H groups (where X=O, N), 
resulting in increased interactions.~\cite{viswamitra1993evidence} 
Therefore, the orientations where hydrogen atoms point towards the pyrene core are preferred and result in higher adsorption energies. 
In the case of NH$_3$ in B$_d$ and H$_2$O in C$_u$ and C$_v$ configurations, 
the adsorption is likely to occur through a polar hydrogen-$\pi$ interaction between the H atoms of NH$_3$ and $\pi$-system of the pyrene. 
We found the lowest adsorption energies for H$_2$O in T$_u$ and NH$_3$ in T$_u$ and C$_u$ configurations. 
Finally, we note that the adsorption of H$_2$O is energetically more favorable with respect to NH$_3$, 
while NO$_2$ has on average the highest adsorption energy 
due to the strong interaction stemming from the open-shell nature of this adsorbate. 
This also results in a significant electron transfer from the pyrene to the adsorbate, 
which we identified from a Bader analysis~\cite{tang2009grid} of the charge density distribution, as reported in Table~\ref{table:Charge}. 

We have also investigated the effect of adsorbing the molecules 
at different positions of the pyrene core. 
We found that this effects on the adsorption energy and the charge transfer 
is much weaker than the orientation of the molecule. 
For instance, we can consider the case of NH$_3$ and H$_2$O adsorbed in the B$_d$ configuration 
on the C-C bond in the center of the meta-substituted pyrene. 
For NH$_3$ we find a similar adsorption energy $E_a=-0.155$~eV 
and a slightly lower charge transfer $\Delta Q=-0.004~|e|$, 
while for H$_2$O we find $E_a=-0.194$~eV and $\Delta Q=-0.022~|e|$, 
which are similar to the results obtained for the C$_v$ (C$_d$) configuration 
(see Tables~\ref{table:Ad}~and~\ref{table:Charge}).  
We expect similar results when considering other adsorption positions, 
and similar conclusions are also drawn in the literature 
for the adsorption of several molecules on graphene.~\cite{lin2013adsorption}

\subsection*{Chemiresistive sensing properties}
The idea behind a chemiresistor is that its electronic transport properties are modulated in response to variations of the nearby chemical environment. 
The underlying mechanism is highly dependent on the molecule-surface interaction.  
In the following, we demonstrate that the setup with a pyrene molecular junction allows to detect the presence of individual physisorbed molecules, 
and it can therefore be used for gas sensing even at very low concentrations.  

The most significant parameters for the characterization of the sensing performance are the sensitivity and the chemical selectivity, 
which we will assess by an analysis of the electron transmission properties for each adsorbate configuration. 
The sensitivity for a given adsorbate is defined as 
\begin{equation}
    {S(\%)} = \bigg[ \frac{T(E_F) - T_{\mathrm{0}}(E_F)} {T_{\mathrm{0}}(E_F)} \bigg] \times 100\%,
\end{equation}
where $T$ and $T_{\mathrm{0}}$ are the transmission functions evaluated at the Fermi level $E_F$ of the junction with and without the adsorbate, respectively. 
The selectivity is instead related to the ability to identify a specific adsorbate from the others, and therefore relies on differences in the individual sensitivities. 

\begin{figure}[tbp]
    \begin{center}
    \includegraphics[width=0.38\textwidth]{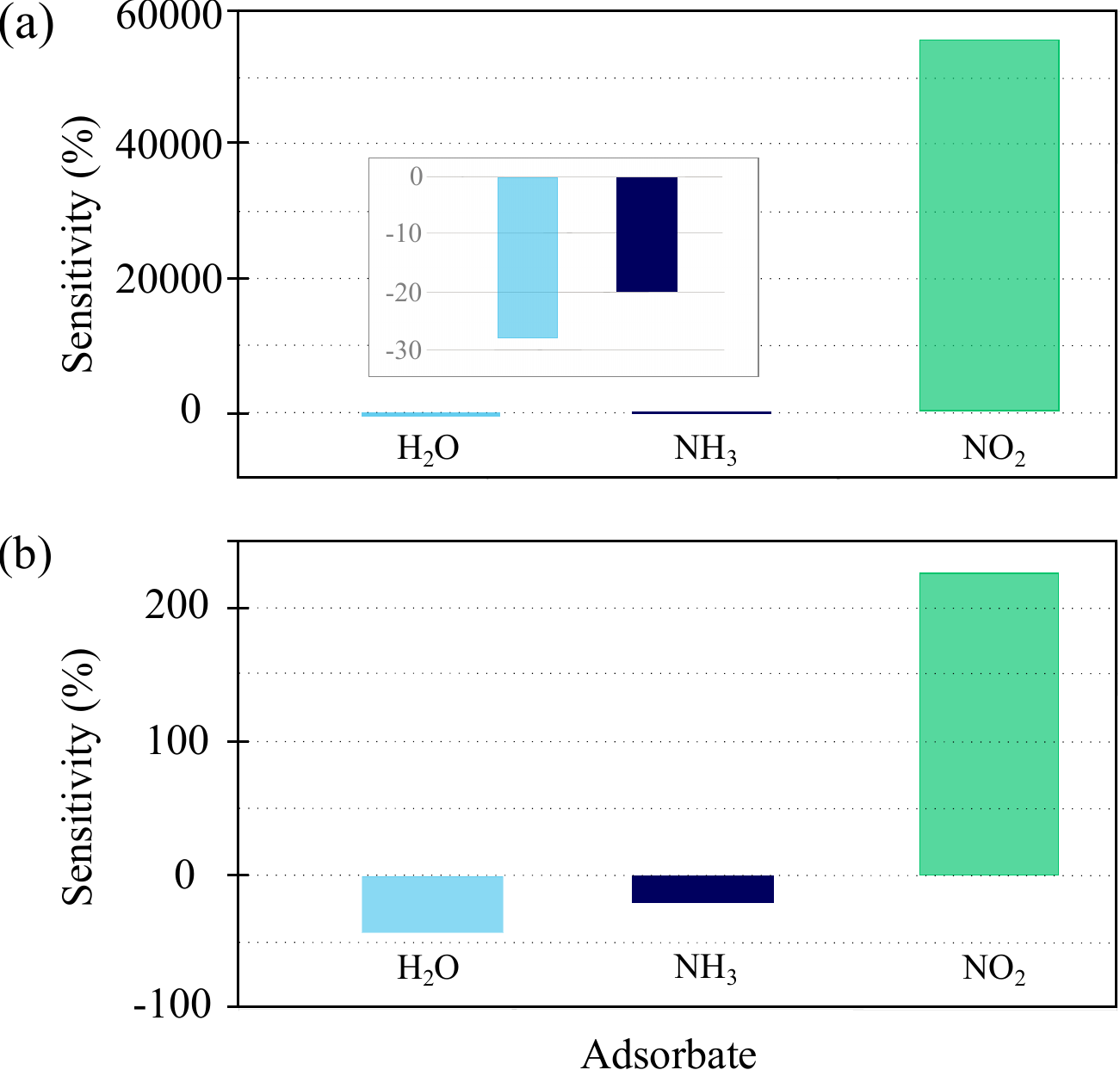}
    \end{center}
    \caption{The calculated sensitivities for H$_2$O, NH$_3$, and NO$_2$ gas molecules adsorbed on the (a) meta- and (b) para-substituted pyrene junctions.  }
    \label{fig:Sensitivity}
\end{figure}

In the following, we focus on the most stable configuration for each adsorbate 
but we stress that all the arguments that we bring forward and our conclusions are consistent 
for all other configurations (not shown). 
We present the results in Fig.~\ref{fig:Sensitivity} where positive and negative 
sensitivities indicate an enhancement or suppression of the electronic conductance. 
Two features clearly stand out, i.e., the sensitivity for NO$_2$ is immensely higher than those of H$_2$O or NH$_3$, 
and also much higher (about two orders of magnitude) in the meta- than in the para-substituted pyrene junction. 
Hence, NO$_2$ is easy to detect because its adsorption triggers a significant chemiresistive response, 
while the signal is weaker and very similar for H$_2$O and NH$_3$, making their sensing and identification harder. 
High sensitivity was already experimentally reported for adsorption of NO$_2$ on graphene.~\cite{schedin2007detection}

\begin{figure}[tbp]
    \begin{center}
     \includegraphics[width=0.49\textwidth]{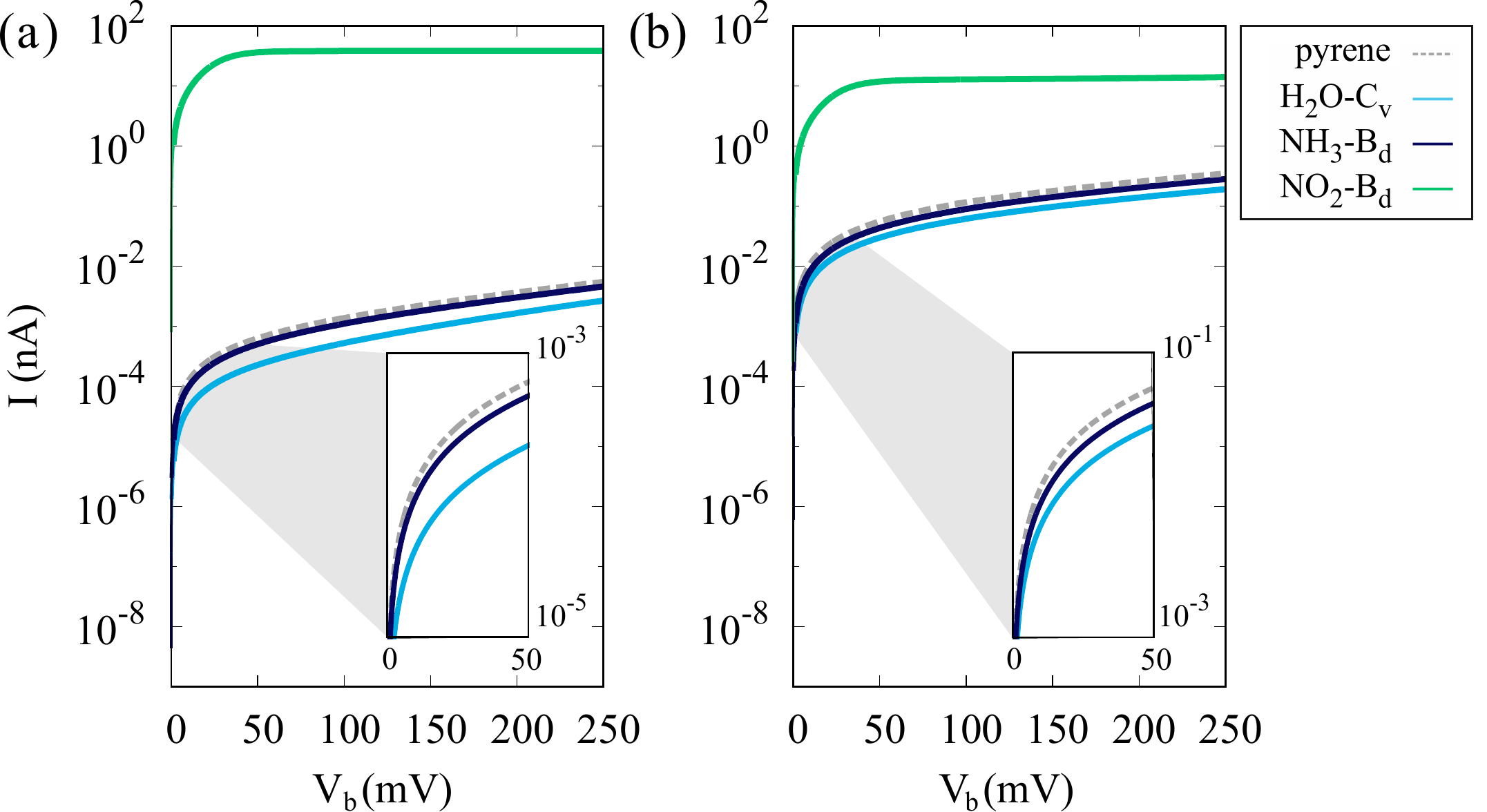}
    \end{center}
    \caption{I-V characteristics before and after the adsorption of H$_2$O, NH$_3$, or NO$_2$ on the (a) meta- and (b) para-substituted pyrene. 
    The grey dashed lines represent the I-V characteristics of junctions without the adsorbate. 
    Each inset highlights the relative positions of the I-V characteristics before and after the adsorption of the adsorbates; H$_2$O and NH$_3$.}
    \label{fig:IV}
\end{figure}

\begin{figure*}[thpb]
    \centering
    \includegraphics[width=0.90\textwidth]{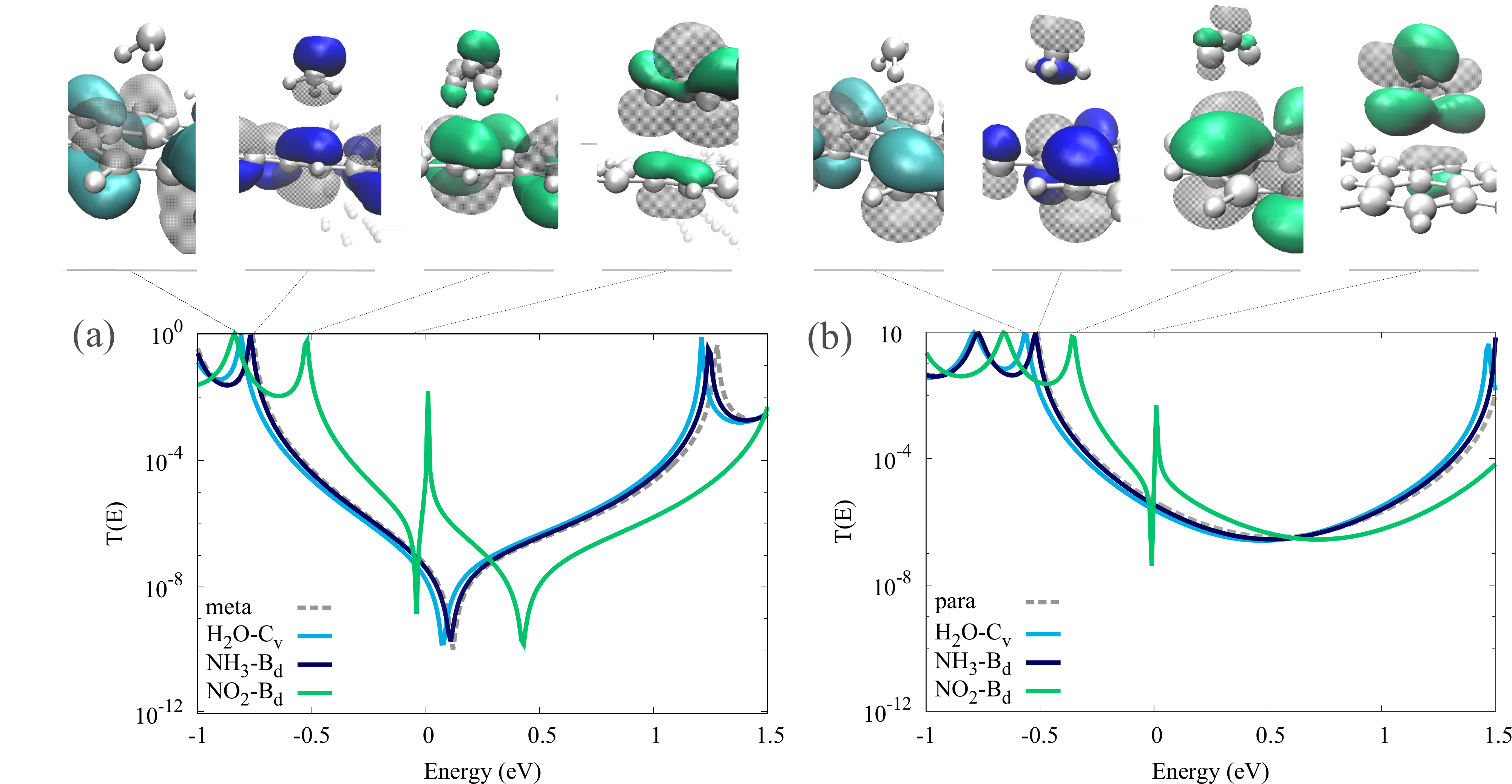}
    \caption{Transmission functions for pyrene junctions in the (a) meta- and (b) para-configurations with the adsorbate: H$_2$O, NH$_3$, NO$_2$ (solid lines), 
    and without the adsorbate (grey dashed line). 
    The insets above (from left to right) show the MOs corresponding to the HOMO of the junction for each adsorbate, 
    and in addition the MO localized at the NO$_2$ giving rise to the Fano resonance.}
    \label{fig:TF}
\end{figure*}

We now look at the current-voltage characteristics, which provide information about the difference in electronic structure of the junctions. 
The current ($I$) versus applied bias voltage ($V_b$) is shown in Fig.~\ref{fig:IV} 
for both the meta- and para-connected junctions before and after the adsorption of the molecules. 
We observe a significant enhancement of the electric current in the presence of NO$_2$, 
which saturates for $V_b>25$~mV. 
This is the consequence of an asymmetric Fano resonance as we will demonstrate later. 
The adsorption of H$_2$O and NH$_3$ results in a suppression of the electric current 
but the I-V curves are featureless since the considered bias window is well inside the junctions HOMO-LUMO gap, 
i.e., $eV_b \ll \Delta \approx 2$~eV. 
This means that, even at a finite bias, it remains more difficult to detect either H$_2$O or NH$_3$, or distinguish between them, in contrast to NO$_2$. 
We shall see further below that this limitation can be overcome by applying a gate voltage.

\subsection*{Role of QI effects}
The difference in the electronic response of the pyrene junction to the physisorption can be completely understood in terms of quantum interference (QI) effects, with different origins.  
We observe, by comparing panels (a) and (b) in Fig.~\ref{fig:IV}, that the bias-driven electronic current is generally higher for junctions with a para-substituted pyrene. 
This is a consequence of the QI properties of the pyrene core,~\cite{sangtarash2015searching,sengul2021electrode} 
which exhibits CQI or DQI in the para- and meta-configuration, respectively. 
While one could expect CQI to be beneficial for sensing applications because of the higher electric signal, 
it is worth noting that junctions exhibiting DQI are found to be more responsive to changes in the chemical environment (see Fig.~\ref{fig:Sensitivity}). 

The role played by QI effects becomes evident by analyzing the electronic transmission functions 
for the junctions in the most stable adsorption configurations, 
which are shown in Fig.~\ref{fig:TF}(a) for meta-substituted and in Fig.~\ref{fig:TF}(b) for para-substituted pyrene, 
where energies are given with respect to the Fermi level $E_F$. 
There are three key features in the electronic transmission, that we shall discuss in detail: 
(i) all transmission functions in the meta configuration exhibit an antiresonance within the HOMO-LUMO gap as hallmark of DQI, 
which is absent in the para configuration,~\cite{sengul2021electrode} 
(ii) the transmission functions of the junctions with H$_2$O or NH$_3$ adsorbates are very similar 
before (grey dashed lines) and after (color solid line) the adsorption, 
although the position of the frontier MOs and the antiresonances with respect to the Fermi level are different; 
(iii) the presence of NO$_2$ instead results in a sizable shift of the Fermi level 
towards the HOMO, as a consequence of the charge transfer from the pyrene molecule to the adsorbate, 
and it also gives rise to an asymmetric resonance at the Fermi level. 

In the junctions exhibiting DQI, the electronic transmission function is strongly suppressed in an energy window around the antiresonance. 
It is known that the energy $\epsilon^{\mathrm{DQI}}$ of the antiresonance depends 
on a cancellation in the electronic Green's function involving all MOs,~\cite{zhao2017destructive,sengul2021electrode,valli2018quantum,valli2019interplay} 
and therefore, the \textit{relative} position of the antiresonance with respect to frontier MO resonances and the Fermi level 
is not necessarily the same in junctions with and without the adsorbate, since the latter can couple differently with each MO. 
In particular, in Fig.~\ref{fig:TF}(a,b), we observe a shift in the Fermi level alignment due to the presence of the adsorbate. 
The hierarchy of the shifts is found to be proportional to the charge transfer between the adsorbates 
and the junction,~\cite{stadler2006fermi, stadler2007fermi,stadler2010conformation,kastlunger2013charge,schwarz2016charge} 
which we found to be weak, i.e., $\Delta Q \sim 0.01 |e|$ for NH$_3$ and $ \Delta Q\sim 0.02 |e|$ for H$_2$O 
(see Table~\ref{table:Charge} for the values in all configurations). 
However, in the meta configuration, after the adsorption, the energy of the antiresonance $\epsilon^{\mathrm{DQI}}$ lies \textit{closer} to the Fermi level. 
This results in a change of the zero-bias conductance $G = G_0 T(E_F)$ (in unit of the conductance quantum $G_0=e^2/h$) 
from $G = 4.9 \times 10^{-8} \ G_0$ (value of the junction without adsorbate) 
to $G_{\mathrm{NH_3}} = 3.9 \times 10^{-8} \ G_0$ and $G_{\mathrm{H_2O}} = 1.7 \times 10^{-8} \ G_0$ 
after the adsorption of the corresponding molecule. 
For reference, the corresponding conductance values in the para configuration (i.e., without DQI) are 
$G = 4.5 \times 10^{-6} \ G_0$, $G_{\mathrm{NH_3}} = 3.8 \times 10^{-6} \ G_0$, and $G_{\mathrm{H_2O}} = 2.4 \times 10^{-6} \ G_0$. 
This shows that the steep variation of the transmission function around the antiresonance 
results in a stronger electric response of the junction even to a relatively weak changes of the Fermi level alignment. 

For NO$_2$, it is a completely different story. As mentioned above, after the adsorption of NO$_2$, 
the transmission function exhibits 
a sharp asymmetric resonance at the Fermi level,~\cite{schedin2007detection,leenaerts2008adsorption} 
which is the reason the sensitivity (as well as the bias-driven current) is significantly higher than for the other molecules.  
It is worth noting that NO$_2$ is a relatively strong electron-withdrawing molecule,~\cite{wehling2008molecular} 
as can be confirmed by the significant electron transfer of $\Delta Q \approx 0.2 |e|$ (see Table~\ref{table:Charge}).
The weak coupling between this localized MO at NO$_2$ and the delocalized $\pi$ system defining the pyrene conducting backbone gives rise to a Fano resonance at the Fermi level.~\cite{fano1961effects,nozaki2013prediction,stadler2005forces,stadler2009quantum,stadler2010conformation}

The combination of the two effects, i.e., the Fermi level being closer to the HOMO resonance and the Fano resonance, 
results in the enhanced electric current observed in the I-V characteristics of the junction in both the meta- and the para-configurations (see Fig.~\ref{fig:IV}). 
This clarifies the origin of the significant difference in sensitivity between NO$_2$ and the other adsorbates. 
The above rationale is confirmed by a MO analysis. In the insets of Fig.~\ref{fig:TF}, 
we show (from left to right) the HOMO of the junction in the presence of H$_2$O, NH$_3$, and NO$_2$, 
as well as a MO localized mostly on the NO$_2$ molecule, which is responsible for the emergence of the Fano resonance. 

The above analysis is general, and in particular it also applies to all other adsorption configurations listed in Table~\ref{table:Ad}. 
However, since the position of the antiresonance is sensitive to the detailed adsorption configuration the resulting sensitivities can fluctuate 
around the values reported in Fig.~\ref{fig:Sensitivity}.

\section*{QI-enhanced sensitivity}

\begin{figure}[btp]
    \begin{center}
        \includegraphics[width=0.5\textwidth]{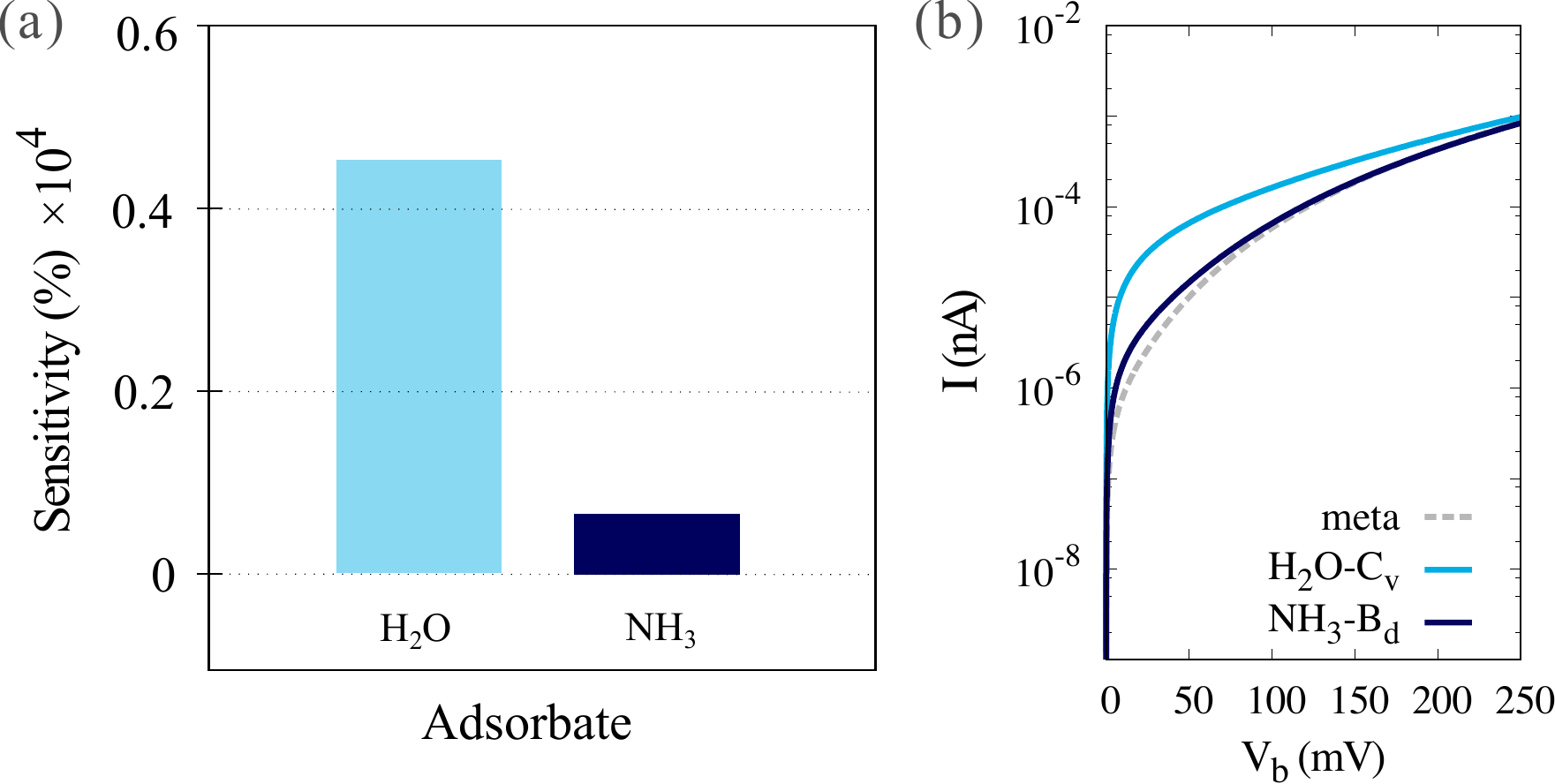}
    \end{center}
    \caption{Chemiresistive response for \textit{calibrated} meta-substituted pyrene sensor, 
    where a gate voltage $V_g=120$~mV is applied to align the antiresonance to the Fermi level and enhance the effect of DQI. 
    (a) calculated sensitivities for H$_2$O, NH$_3$ adsorbates, 
    (b) I-V characteristics before (grey dashed line) and after (color solid lines) the adsorption.}
    \label{fig:IV-meta}
\end{figure}

Due to the steep suppression of the electronic transmission function in the vicinity of an antiresonance, 
junctions exhibiting DQI are expected to display a stronger chemiresistive response, triggered by the adsorption, 
which should allow distinguishing between the signals attributed to individual molecules.~\cite{prasongkit2016quantum} 
However, the sensor performance to detect molecules such as H$_2$O and NH$_3$, 
is still comparable for junctions including meta- and para-substituted pyrene (see Fig.~\ref{fig:Sensitivity}). 
In the following, we propose a general protocol to enhance the performance of single-molecule sensors by exploiting DQI. 
The key idea is to \textit{calibrate} the sensor by applying a gate voltage $V_g$ 
in order to tune the position of the DQI antiresonance of the junction (in the absence of adsorbates) to the Fermi level. 
Reduced screening in organic molecules or two-dimensional materials (such as graphene) 
allow to sweep $V_g$ across the HOMO-LUMO gap to find the configuration with the highest resistance, defined as 
\begin{equation}
    G^{*} = \frac{e^2}{h} \min_{V_g} \big[ T(E_F-eV_g) \big]. 
\end{equation}
In our numerical simulations, 
we identify the value of G$^*$ at $eV_g = E_F - \epsilon^{\mathrm{DQI}} = 120$~meV. 
At that gate voltage, any variation of the chemical environment causing a shift in the position of the antiresonance will trigger the highest possible change in conductance. 
The corresponding values are $G = 1.0 \times 10^{-10} \ G_0$ for the junction without adsorbates, 
$G_{\mathrm{NH_3}} = 5.2 \times 10^{-10} \ G_0$, and $G_{\mathrm{H_2O}} = 4.9 \times 10^{-9} \ G_0$ 
after the adsorption of the corresponding molecule. 
Hence, the calibration process results not only in a higher sensing signal overall for all molecular species considered, 
but it also enhances the difference in the chemiresistive response for different adsorbates 
which is now one order of magnitude comparing NH$_3$ and H$_2$O. 
In Fig.~\ref{fig:IV-meta}, we report both the sensitivity data and the corresponding I-V characteristics. 
In terms of sensitivity, the chemiresistive response is $\sim 5000\%$ for H$_2$O and $\sim 500\%$ for NH$_3$, 
compared to the values of $\sim 30\%$ and $\sim 20\%$, respectively, obtained before the calibration with the gate voltage (see Fig.~\ref{fig:Sensitivity}).  
At low-bias, there is a clear difference (over an order of magnitude) between the electric currents registered for H$_2$O and NH$_3$. 
The difference disappears at larger bias as the contributions from the transmission at energies further away from the antiresonance become dominant and hide the DQI effects. 
Remarkably, the calibration process also removes the dependence 
of the results from the position of the antiresonance of the device, 
so that the sensitivity only arise from the changes of the chemical environment 
induced by the presence of the adsorbate. 

Since NO$_2$ gives rise to a Fano resonance pinned at the Fermi level, the calibration process we reported above is not very relevant for this adsorbate.

\section*{Conclusions}
We presented a detailed analysis of the chemiresistive response of a pyrene single-molecule junction 
to the adsorption of individual gaseous molecules, aiming at a characterization of its performance as a chemical gas sensor. 
We studied the intermolecular interactions of the pyrene core with H$_2$O, NH$_3$, and NO$_2$, 
by varying the adsorption sites and adsorbate orientations, 
focusing on the changes in the electronic transport properties induced by the presence of the adsorbates. 
In particular, we highlight the pivotal role which can be played by DQI effects in the sensing process. 

The adsorption of NO$_2$ yields 
the strongest chemiresistive response 
among the molecules investigated here. Indeed, due to its strong electron-withdrawing character,
a partially filled molecular state of NO$_2$ is pinned to the Fermi level and it couples with the electronic states of the junction, 
giving rise to a sharp asymmetric Fano resonance. 
This mechanism is responsible for the strong electronic signal observed for NO$_2$. 

In contrast, both H$_2$O and NH$_3$ do not provide a localized state close to E$_F$ which is necessary for a Fano resonance, 
and result in relatively lower sensitivity and poor chemical selectivity between those two. 
This issue can be overcome considering that meta-substituted pyrene molecular junctions exhibit DQI, 
whose hallmark is a steep suppression of the electronic transmission due to the presence of an antiresonance. 
The position of the antiresonance relative to the Fermi level is very sensitive to changes in the chemical environment 
and therefore it strongly influences the chemiresistive response.  
Finally, we propose a protocol for the calibration of the QI-sensor device. 
The application of a gate voltage allows to tune the junction (in the absence of any adsorbate) 
in the configuration of maximal resistance, 
which is realized when the antiresonance is aligned to the Fermi level. 
This is shown to boost significantly, i.e., by one or two orders of magnitude, 
the sensing performance in terms of both sensitivity and selectivity of individual molecules. 
The calibration process optimizes the sensing performance of the device 
by disentangling the chemiresistive response 
from intrinsic properties of the device, such as the position of the DQI antiresonance. 
This allows to better characterize the effects induced by the adsorbate. 


In conclusion, we demonstrated that the adsorption of individual gas molecules could be detected 
in pyrene-based single-molecule junctions, and we proposed QI as a paradigmatic mechanism 
to enhance the sensing performance of nanosensors based on graphene functional units. 
Our work is significant for both physical and biochemical applications, 
and it is encouraging for the prospects of the technological improvement of graphene-based gas sensors. 


\section*{Acknowledgements}
We thank S.~Tkaczyk for valuable discussions. 
We acknowledge financial support from the Austrian Science Fund (FWF) through project P 31631. 
Calculations have been performed on the Vienna Scientific Cluster (VSC), project No. 71279.

\bibliographystyle{apsrev}
\bibliography{arXiv2}

\end{document}